# Spin-phonon coupling in multiferroic $R$CrO$_3$ ($R$-Y, Lu, Gd, Eu, Sm): A Raman study

Venkata Srinu Bhadram, B. Rajeswaran, A. Sundaresan and Chandrabhas Narayana

*Chemistry and Physics of Materials Unit, Jawaharlal Nehru Centre for Advanced Scientific Research, Jakkur P. O., Bangalore 560 064, India.*

**Abstract** – Raman study on a select few orthochromites, $R$CrO$_3$ ($R$ = Y, Lu, Gd, Eu and Sm) shows that the phonon behavior at T$_N$ in compounds with magnetic $R$-ion (Gd and Sm) is remarkably different from that of non-magnetic $R$-ion (Y, Lu and Eu). While anomalies in most of the observed phonon frequencies in all these compounds may result from the distortion of CrO$_6$ octahedra due to size effect and magnetostriction arising from Cr-ordering, the anomalous behavior of their linewidths observed at T$_N$ for the compounds with only magnetic $R$-ion suggests spin-phonon coupling. The presence of spin-phonon coupling and the anomalies in the low frequency modes related to $R$-ion motion in orthochromites ($R$ = Gd and Sm) support the suggestion that the coupling between 4f-3d moments play important role in inducing switchable electric polarization.

**Introduction.**–Magnetoelectric multiferroic materials with their coupled ferroelectric and ferromagnetic order parameters are promising for developing a new generation of both electrically and magnetically controlled multifunctional devices [1-5]. Multiferroic materials are broadly classified into two types; in type I multiferroics, the ferroelectricity and magnetism occur at high temperatures but with different temperature scale. However, the coupling between the two order parameters are rather weak. Type II multiferroics are generally centrosymmetric and magnetic where the ferroelectricity is caused by certain type of magnetic ordering. In the well-known example of TbMnO$_3$, the manganese moments order antiferromagnetically at T$_N$ = 41 K and at 25 K it undergoes another magnetic transition below which a cycloidal spin structure breaks the inversion symmetry and thus induces ferroelectricity [6]. In case of HoMnO$_3$[7], a collinear magnetic ordering with E-type magnetic structure gives rise to ferroelectricity. It has been shown recently that canted antiferromagnetic ordering with two non-equivalent spin pairs in the orthoferrite, SmFeO$_3$[8] induces ferroelectric polarization at the magnetic ordering temperature of iron. Rajeswaran *et al.*[9] have reported electric polar order at the magnetic ordering temperature of chromium in the isostructural orthochromites, $R$CrO$_3$ with magnetic $R$ ion, where the interactions between $R^{3+}$ and Cr$^{3+}$ ions have been suggested to be responsible for the switchable polarization. In such systems, the interplay between magnetic and electric order could be mediated through spin-lattice coupling. Thus, probing the local structure would provide a good understanding of the multiferroic behavior in this class of materials.

Raman spectroscopy is an ideal technique to study the local structural changes due to magnetic ordering effects in materials. Recent Raman reports exemplify the ability of Raman to elucidate the physics of multiferroics [10-12]. Raman spectroscopy of rare-earth orthorhombic manganites, $R$MnO$_3$ ($R$ = Pr, Nd, Sm, Eu, Gd, Tb, Dy, Ho, Y) have been studied extensively in the past [13-18]. In case of hexagonal YMnO$_3$[19] and LuMnO$_3$[20], ferroelectric to paraelectric phase transition is accompanied by tilting of MnO$_5$ polyhedra which is correlated with the changes in Raman modes related to rare-earth atom displacements. In the case of orthorhombic TbMnO$_3$, new Raman modes have been observed at the cycloidal spin ordering temperature suggesting their magnetoelectric origin [21]. Though orthorhombic manganites have been studied extensively in the past, very few reports of Raman studies exist in the case of orthochromites.

In the present work, we report the temperature-dependent Raman studies of rare-earth orthochromites, $R$CrO$_3$ with magnetic (Gd and Sm) and nonmagnetic (Y, Lu and Eu) $R$- ions to exemplify the role of rare-earth magnetism in inducing the ferroelectricity in these materials. We have chosen these rare-earth ions to account for the effect of both rare-earth ionic size and its magnetism on the spin-phonon coupling and ferroelectricity in $R$CrO$_3$. Lu$^{3+}$ and Y$^{3+}$ are both nonmagnetic which have smaller and comparable ionic radii, whereas nonmagnetic Eu$^{3+}$ has higher ionic radii and is comparable to the size of the magnetic Gd$^{3+}$ and Sm$^{3+}$ ions. Such careful selection is indeed a necessary prerequisite to demarcate the effects that can arise specifically due to either the magnetism of $R^{3+}$ ion or the size of $R^{3+}$ ion or both. Due to leakage effects and oxygen nonstoichiometry in LaCrO$_3$, we have not chosen it in our study even though La$^{3+}$ is nonmagnetic.

The observed anomalies in the line-widths of the phonon modes in GdCrO$_3$ and SmCrO$_3$ indicate spin-phonon coupling in these systems. In contrast, the line-width of orthochromites, YCrO$_3$, LuCrO$_3$ and EuCrO$_3$ with nonmagnetic rare-earth ion do not show any anomaly at $T_N$ which indicates that Cr ordering alone is not responsible for spin-phonon coupling. Moreover, the low frequency lattice modes related to $R$ atom vibrations undergo anomalous changes below $T_N$ in SmCrO$_3$ and GdCrO$_3$ along with significant changes in linewidth indicating phonon mediated spin interactions between $R^{3+}$ and Cr$^{3+}$ ions resulting in a energy minimized state suggesting $R$-ion involved ferroelectric polarization. Such changes are not observed in the case of non-magnetic rare-earth orthochromites.

**Experimental.**–The details of sample preparation and characterization are reported elsewhere [9]. Raman spectroscopic measurements have been carried out in back scattering geometry using 532 nm excitation wavelength and a custom built spectrometer equipped with a SPEX TRIAX 550 monochromator and a liquid nitrogen cooled charge-coupled device (CCD; Spectrum One with CCD 3000 controller, ISA JobinYovn) [22]. Laser power at the sample was kept at ~6mW and typical spectral acquisition time is 1 min. Temperature dependent measurements were done using heating stage unit (Linkam THMS 600) equipped with a temperature-controller (Linkam TMS 94) with temperature accuracy of ±1K. Lorentzian functions were used to fit the spectral profile.

**Results and discussion.**– Fig. 1 shows the room temperature Raman spectra of $R$CrO$_3$ ($R$=Y, Lu, Gd, Eu and Sm) in the spectral range 100-600 cm$^{-1}$. $R$CrO$_3$ has an orthorhombic (P$nma$) structure with 24 Raman active modes which are classified as $\Gamma_{Raman}$= 7A$_g$+ 5B$_{1g}$+7B$_{2g}$+5B$_{3g}$ [15].These modes are assigned by following the earlier Raman reports on orthorhombic $R$MnO$_3$ and the recent room-temperature Raman study on $R$CrO$_3$ systems [14, 15, 24]. The phonon modes below 250 cm$^{-1}$ in YCrO$_3$ and below 200 cm$^{-1}$ in all the other samples are related to lattice modes involving $R$ atom vibrations. Region above 200 cm$^{-1}$ (250 cm$^{-1}$ in the case of YCrO$_3$) consists of various modes involving vibrations of $R$ atom and oxygen. A$_g$(1) mode is related to the anti-symmetric stretching vibrations of CrO$_6$ octahedra. B$_{1g}$(3), A$_g$(2) are octahedral rotations around crystallographic y-axis and B$_{1g}$(4), A$_g$(4) are rotations around x-axis (P$nma$ setting). The doublet (A$_g$(7) and B$_{2g}$(7)) at around 300 cm$^{-1}$ in YCrO$_3$, LuCrO$_3$, GdCrO$_3$, EuCrO$_3$ and a singlet (A$_g$(7)) in SmCrO$_3$ are related to $R$-O vibrations. B$_{3g}$(3) is bending mode of CrO$_6$ octahedra. Most of these phonon modes shift to low frequency with increase in ionic radii of $R$-ion ($r_Y$>$r_{Lu}$>$r_{Gd}$>$r_{Eu}$>$r_{Sm}$) as seen in the case of orthorhombic manganites [14]. A detailed spectral analysis of these systems has been reported recently [23].

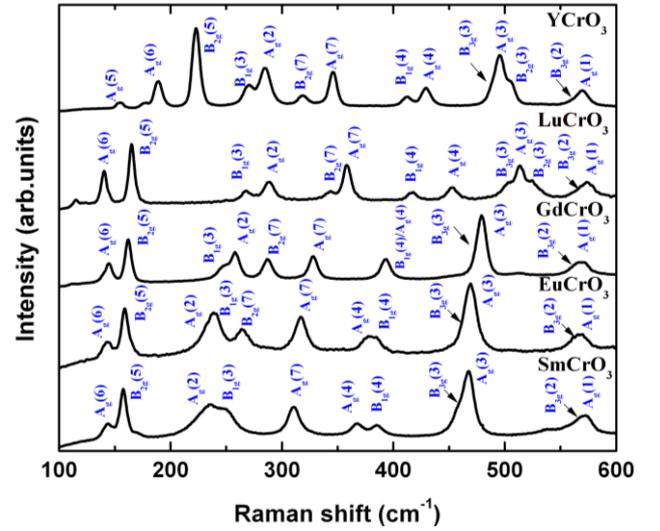

Fig. 1:(Color online) Unpolarized Raman spectra of $R$CrO$_3$($R$= Y, Lu, Gd, Eu and Sm) at 298 K. Mode assignments have been done based on Ref. [13, 14, 23].

The temperature effects on the anti-symmetric stretching mode A$_g$(1) frequency and corresponding line widths are plotted in fig. 2 in the temperature range 77-298K. Temperature dependent behavior of phonon mode of frequency ω is given as [24]

$$\omega(T) = \omega(0) + \Delta\omega_{qh}(T) + \Delta\omega_{anh}(T) + \Delta\omega_{sp-lat}(T) + \Delta\omega_{el-ph}(T)$$

where ω(0) is the frequency at zero K. The term $\Delta\omega_{qh}(T)$ includes the changes in phonon frequency due to the quasi-harmonic effect which corresponds to the change in the lattice volume. Whereas $\Delta\omega_{anh}(T)$ gives the intrinsic anharmonic contribution to the frequency. $\Delta\omega_{sp-lat}(T)$ represents the change in phonon frequency due to spin-lattice (spin-phonon) coupling which is caused by the phonon modulation of spin-exchange integral. Effects of electron-phonon coupling on phonon frequency is given by $\Delta\omega_{el-ph}(T)$ which can be neglected when the carrier concentration is low.

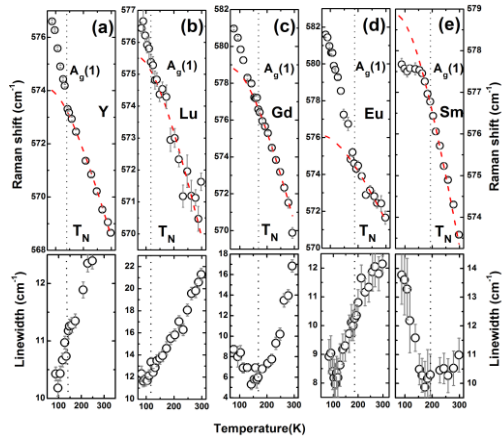

Fig. 2:(Color online) (Top panel) Temperature dependence of the $A_g(1)$ mode frequencies of (a) $YCrO_3$ (b) $LuCrO_3$ (c) $GdCrO_3$ (d) $EuCrO_3$ (e) $SmcrO_3$ in the temperature range 77-298K. Red dotted lines represent the anharmonic contribution to the phonon frequency as mentioned in the text. (Bottom panel) corresponding linewidths.

As seen in fig. 2, phonon frequencies increase with decreasing temperature up to $T_N$ in all three systems which is attributed to the anharmonic effect. The dotted line in the fig. 2 is the fit to the experimental data above $T_N$ with the equation [25],

$$\omega_{anh} = \omega_0 - C(1 + 2/(e^{\hbar\omega/KT} - 1))$$

where $\omega_0$ and C are adjustable parameters. It is clear from fig. 2 that $A_g(1)$ phonon frequency deviates from the above relation below $T_N$ in all the three systems, which could arise from the quasi-harmonic effects along with spin-phonon coupling. However, it is interesting to note the different behavior of phonons below $T_N$ for different R-ions. An anomalous hardening of the $A_g(1)$ mode is observed in all the samples except in $SmCrO_3$ wherein a softening is observed. The observed hardening of $A_g(1)$ mode is consistent with the earlier observation where a reduction in unit cell volume is reported for $YCrO_3$ and is attributed to magnetostriction [26]. Though we are not aware of such studies in other samples, it is possible that a similar magnetostriction effect can occur in all these samples due to magnetic ordering of $Cr^{3+}$ ions. The observation that the phonon mode below $T_N$ softens in $SmCrO_3$ and hardens in $RCrO_3$ (R=Y, Lu, Gd, Eu) may be related to different magnetic interactions between magnetic $R^{3+}$ and $Cr^{3+}$ ions as reported earlier [9].

To give more insight into the origin of anomalous behavior of $A_g(1)$ mode frequency, we have plotted the temperature dependence of the corresponding linewidth in fig. 2 (bottom panel). Since Raman linewidths are related to phonon lifetime which will not be effected by subtle lattice volume changes due to magnetostriction, it is better to study linewidths rather than phonon frequencies to observe the spin-phonon coupling. We observe that the line width increases below $T_N$ in the case of magnetic R-ion (Gd, Sm) whereas no significant changes are observed in the case of nonmagnetic R-ion (Y, Lu and Eu). Indeed, Raman linewidths are expected to decrease monotonously with temperature due to anharmonicity [25] as seen in the case of nonmagnetic R-ion (see fig. 2).

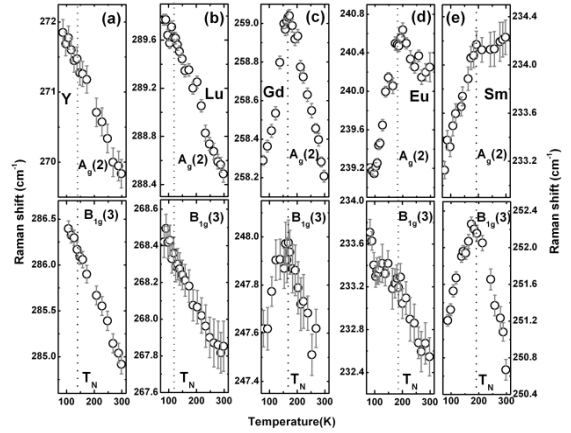

Fig. 3:(Color online) Temperature dependence of in- phase $A_g(2)$ and out-of- phase $B_{1g}(3)$ octahedral rotations with respect to y-axis in (a)$YCrO_3$ (b) $LuCrO_3$ (c) $GdCrO_3$ (d) $EuCrO_3$ (e) $SmCrO_3$.

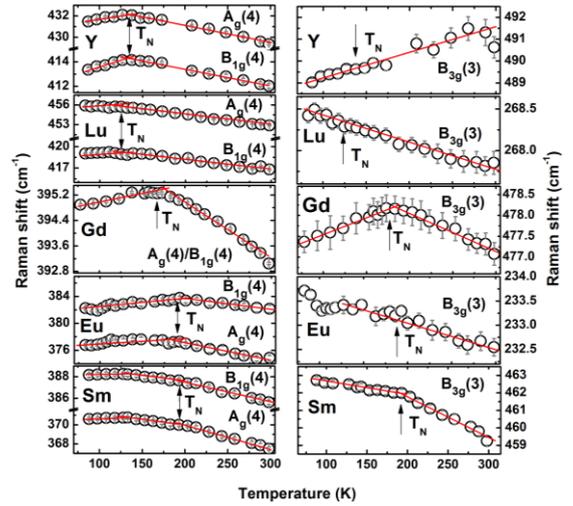

Fig. 4:(Color online) (Left panel) Temperature dependence of in-phase ($A_g(4)$) and out-of-phase ($B_{1g}(4)$) $CrO_6$ rotations with respect to x-axis. (Right panel) Temperature effects on bending mode $B_{3g}(3)$ in $RCrO_3$ (R=Y, Lu, Gd,Eu andSm). Solid lines are guide to the eye.

In contrast, an anomalous increase in the line width below $T_N$ is observed in magnetic R-ion which can be observed from the temperature dependence of the $A_g(1)$ mode profile (as shown in supplementary information[1] (fig. S1) for the case of GdCrO$_3$) signifies the contribution from effects that change the phonon lifetime. Generally, phonon lifetimes are effected by various processes such as spin-phonon coupling [27, 28] and electron-phonon coupling [29]. Since these materials are insulators, we can unequivocally attribute the change in linewidth to spin-phonon coupling.

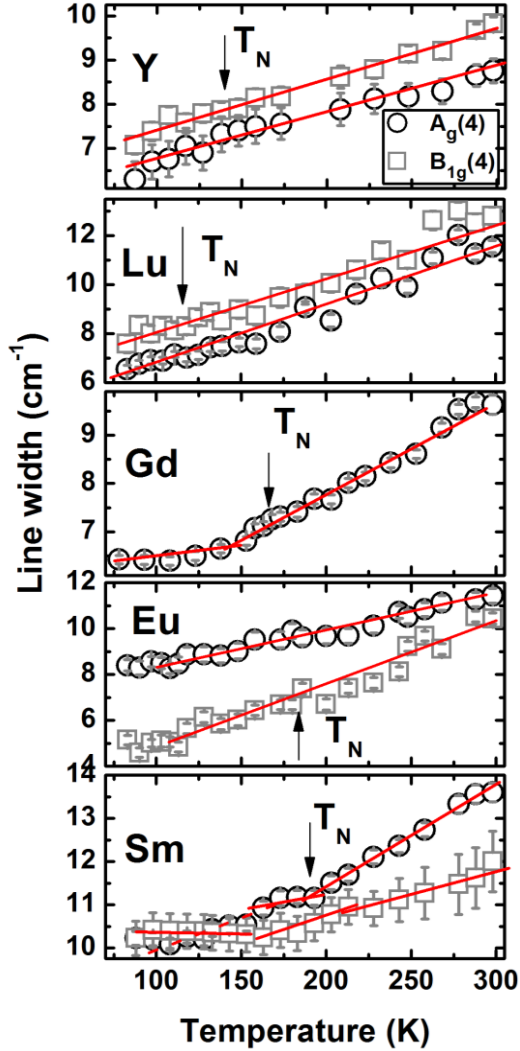

Fig. 5: (Color online) Effect of the temperature on B$_{1g}$(4) and A$_g$(4) phonon line widths in $R$CrO$_3$ ($R$= Y, Lu, Gd, Eu and Sm). Solid lines are guide to the eye.

We now discuss the temperature dependence of the in-phase and out-of-phase octahedral rotations about y-axis in the temperature range 77 – 298 K which is shown in fig. 3. We observed anomalous softening of the phonon frequencies (A$_g$ (2), B$_{1g}$ (3)) below T$_N$ in the case of Gd and Sm. We also observed anomalies in the corresponding linewidths (not shown here) indicating the spin-phonon coupling in these systems. Remarkably, no significant changes are observed in Y and Lu. In the case of Eu, though the A$_g$(2) phonon mode[1] shows softening, the corresponding linewidth (not shown here) remains unaffected. So the possibility of spin-phonon coupling in EuCrO$_3$ is ruled out. The observed softening of A$_g$(2) mode in Gd, Eu and Sm could be related to the localization of $R^{3+}$ cations due to the relative larger ionic sizes allowing free rotation of CrO$_6$ octahedra. Temperature dependence of in-phase and out-of-phase octahedral rotations about x-axis (A$_g$(4), B$_{1g}$(4)) is shown in left panel of fig. 4. Unlike rotation about y-axis, all the samples show softening in rotation about x-axis below T$_N$ which may be predominantly due to the effect of magnetostriction. The behavior of out-of-phase bending mode, B$_{3g}$(3) as shown in right panel of fig. 4 is similar to that observed for rotational modes about y-axis. It should be noticed that the A$_g$(4), B$_{1g}$(4) and B$_{3g}$(3) modes in SmCrO$_3$ exhibit an additional anomaly near 140 K which may be linked to the observed spin-reorientation transition occurring at much lower temperature (around 40 K) [9].

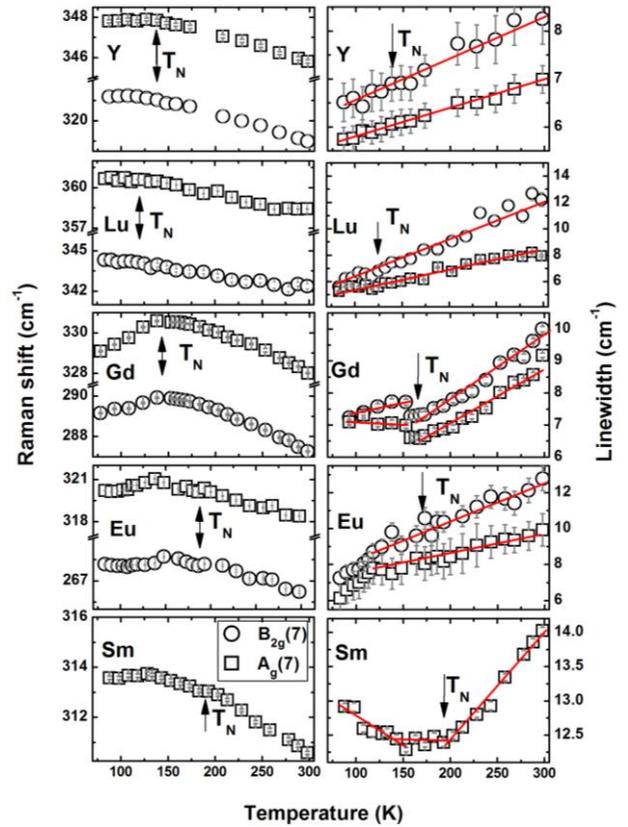

Fig. 6: (Left panel) Temperature dependence of frequencies of A$_g$(7), B$_{2g}$(7) modes involving $R$-O vibrations. (Right panel) Corresponding linewidths. Solid lines are guide to the eye.

---

[1] Supplementary information available at http://desktop.jncasr.ac.in/uploaded/cbhas/Supporting Information for EPL 2013.pdf

To verify whether the anomalous behavior of phonon modes, $A_g(4)$ and $B_{1g}(4)$ is associated with spin-phonon coupling in addition to the contribution from magnetostriction as seen in Y, Lu and Eu, we have analyzed the temperature dependence of linewidth of these modes as shown in fig. 5. It is obvious that only in the cases of Gd and Sm we observe anomalies in linewidth (namely, broadening) around $T_N$ suggesting the presence of spin-phonon coupling. From these observations, we can conclude that spin-phonon coupling is present only in $R$CrO$_3$ with magnetic $R^{3+}$ ions.

Temperature dependence of the $A_g(7)$ and $B_{2g}(7)$ mode frequencies and the corresponding line widths are shown in fig. 6. These are lattice modes involving $R$-O vibrations. We notice from the left panel that all the samples show phonon anomalies at $T_N$ but this is relatively weak in Y, Lu and Eu.
Such a small change in phonon frequency along with no change in phonon line width suggests that these may solely be due to magnetostriction. However, a strong softening of these modes along with the anomalies in their line width (right panel) in Gd and Sm confirms the presence of spin-phonon coupling. The relatively strong softening of these low frequency modes including the $B_{2g}(5)$ mode (fig. 7) further reveals a possible displacement of $R^{3+}$ ions induced by spin-phonon coupling.

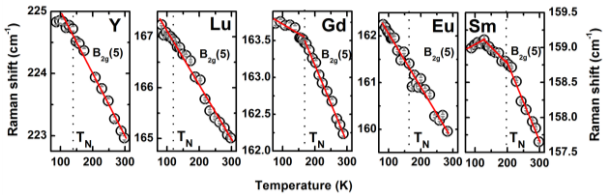

Fig. 7: Temperature dependence of frequency of $B_{2g}(5)$ mode involving pure $R$ atom vibrations in $R$CrO$_3$. Dotted lines are guide to the eye

In the past, efforts were made to understand the correlation between spin-phonon coupling and magnetoelectric effect in various manganites, ferrites and nickelates [10, 30-32]. A theoretical model has been proposed by considering the spin-phonon coupling and Mn spin frustration to explain the origin and nature of the multiferroic phases in $R$MnO$_3$ [33]. It is also suggested that a frustrated intra and inter-plane Mn-Mn, $R$-Mn interactions arise in $R$MnO$_3$ below Mn ordering temperature which would favour the displacement of Mn to minimize the frustration. This displacement of Mn as proved by neutron diffraction studies would lead to ferroelectricity in $R$MnO$_3$ [34]. In the present $R$CrO$_3$ case, the situation might be similar except that the strong spin-phonon coupling, here, is due to the weak ferromagnetic $R$-Cr interactions which may displace the magnetic $R^{3+}$ resulting in ferroelectric polarization [9, 35]. A detailed neutron diffraction studies and first-principle calculations would help in understanding these systems better.

It is to be noted that there are anomalies observed at around 120 K in frequencies as well as linewidths of some phonon modes in the case of EuCrO$_3$ which is consistent with our preliminary ferroelectric studies based on pyroelectric current measurements clearly demonstrate the appearance of polarization in the same temperature regime. However, the understanding of these anomalies observed in Raman studies and polarization measurements in EuCrO$_3$ at a lower temperature (T<$T_N$) is beyond the scope of the present discussion and further studies are planned.

In conclusion, Raman results clearly demonstrate spin-phonon coupling in rare-earth orthochromites $R$CrO$_3$ with magnetic $R$-ion. We observe along with phonon frequency changes, decrease in phonon lifetimes both in modes involving CrO$_6$ octahedra as well as magnetic $R^{3+}$ ion. This clearly suggests a phonon mediated magnetic interactions between $R^{3+}$ and Cr$^{3+}$ ions. These interactions along with the softening of magnetic $R$-ion modes suggest that the origin of ferroelectric polarization in rare-earth orthochromites may be related to the displacement of $R$-ion.

\*\*\*